# CFD Analysis of Latent Heat Energy Storage System with Different Geometric Configurations and Flow Conditions

Pushpendra Kumar Shukla[a*], P. Anil Kishan[a]

[a] School of Engineering, IIT Mandi
* Corresponding author email: D16055@students.iitmandi.ac.in

**ABSTRACT**

The Latent heat storage technology is being used worldwide to bridge the gap between supply and demand of energy. The material store energy during the charging process (melting) and releases energy during the discharging process (solidification). In spite of having various advantages such as high storage energy density, it suffers from the fact that most Phase Change Materials (PCMs) commonly used have a very low thermal conductivity, hence, very slow charging /discharging times. In the current work, a shell and tube type heat exchanger with phase change material on the shell side and heat transfer fluid on the tube side are considered. The effect of flow rate and inlet temperature of heat transfer fluid on melting and solidification times are investigated with single and double pass (counter and parallel) arrangements of Heat Transfer Fluid (HTF). The major difficulty encountered in the melting of the PCM is the accumulation of solid (unmelted) part at the bottom during the charging process, while the liquid part remains at the top during the discharging process, which decreases the efficiency of the system to quite a great extent. In this study, an attempt has been made to improve the efficiency of the system by considering two configurations (double and triple tube) of the shell and tube heat exchanger and it is found that the latter case has better performance.

**NOMENCLATURE**

| | |
|---|---|
| $\alpha$ | Volume thermal expansivity (1/K) |
| $\beta$ | Local liquid fraction (-) |
| g | Gravitational acceleration (m/s$^2$) |
| $T_m$ | Melting point of the PCM (K) |
| $\Delta H$ | Fusion latent heat (kJ/kg) |

**INTRODUCTION**

The energy coming from renewable sources, such as solar, is abundant but intermittent in nature. To use these renewable energy resources, it is crucial to develop efficient energy storage systems that can store energy available in times of surplus supply and provide it in times of high demand and shortage in supply [1]. Conventional solar thermal to hot water systems are utilized for residential applications, with relatively low efficiency and limited utility, particularly at daytime. Storing the energy, using PCM is one of the alternatives, to store thermal energy [2].

Latent heat thermal storage systems using PCMs have been proved to be more efficient than sensible heat storage systems [2]. Heat is added to an energy storage system with the help of HTF, circulating from the solar collectors or other means through the PCM tank. Water is commonly used as HTF. Hot HTF transfers energy to the PCM and melts the PCM and stores thermal energy in the form of latent heat. During the utilization, the HTF extracts the latent heat from PCM. PCMs can be broken down into three categories based on their chemical composition: organic, inorganic, and liquid metals [2]. Each type of PCM is suited for different applications having different melting temperature requirements, ranging from sub-zero centigrade to over 1000 °C. In the current study, we are working with melting points close to 42 °C

Different designs of phase change energy storage systems have been studied. A tube-and-shell is one of the simplest designs and it is most commonly used [3]. Kazemi et al. worked on different fins configuration to improvement the performance of the systems [4]. Various factors must be considered during the design process. The charging-discharging times play an important role in residential applications. Large charging time and discharging times are highly undesirable. Study of charging and discharging times has been done for higher temperature ranges (350-650 °C) like solar thermal power plant applications [3].

In this paper, simulation studies with the conditions prevailing to the energy storage required for space heating were presented. The effects of HTF inlet temperature and flow rate on the charging and discharging times were analysed for both configurations hence a comparison is done for the same amount of PCM. Also, the melting and solidification characteristics of the PCM are examined with variable HTF flow rate and inlet HTF temperature.

## GOVERNING EQUATIONS

The governing equations solved for the simulation of melting and solidifications include conservation equations for mass, momentum, and energy. In this study, the enthalpy-porosity model is adopted, which is widely applied for modeling the PCM melting and solidification processes [5].

The enthalpy-porosity model in 2D can be described as below:

- Continuity Equation

$$\frac{\partial \rho}{\partial t} + \frac{\partial (\rho u)}{\partial x} + \frac{\partial (\rho v)}{\partial y} = 0 \qquad (1)$$

- Momentum Equations

$$\frac{\partial (\rho u)}{\partial t} + \frac{\partial (\rho u u)}{\partial x} + \frac{\partial (\rho u v)}{\partial y} = -\frac{\partial P}{\partial x} + \frac{\partial}{\partial x}\left(\mu \frac{\partial u}{\partial x}\right) + \frac{\partial}{\partial y}\left(\mu \frac{\partial u}{\partial y}\right) + Au \qquad (2)$$

$$\frac{\partial (\rho v)}{\partial t} + \frac{\partial (\rho v u)}{\partial x} + \frac{\partial (\rho v v)}{\partial y} = -\frac{\partial P}{\partial y} + \frac{\partial}{\partial x}\left(\mu \frac{\partial v}{\partial x}\right) + \frac{\partial}{\partial y}\left(\mu \frac{\partial v}{\partial y}\right) + Av$$

$$+\rho g \alpha (T - T_m) \qquad (3)$$

Where $\rho$, $\mu$, $\alpha$, g and $T_m$ are density, dynamic viscosity, volume thermal expansivity, gravitational acceleration and melting point of the PCM, respectively. Au and Av are momentum dissipation source items, which are used for suppressing velocity in the solid and mushy region. A is "porosity function" and is defined by Brent et al. [6] as:

$$A = -C \frac{(1-\beta)^2}{\beta^2 + \varepsilon} \qquad (4)$$

Where $C = 1.0 \times 10^5$ is a mushy zone constant to reflect the melting front morphology, $\beta$ is local liquid fraction (the ratio of liquid PCM volume to the total volume of computational cell), and $\varepsilon = 0.001$ is a small number to prevent division by zero [5].

- Energy Equation

$$\frac{\partial (\rho h)}{\partial t} + \frac{\partial (\rho u h)}{\partial x} + \frac{\partial (\rho v h)}{\partial y} = \frac{\partial}{\partial x}\left(k \frac{\partial T}{\partial x}\right) + \frac{\partial}{\partial y}\left(k \frac{\partial T}{\partial y}\right) \qquad (5)$$

$$h = \begin{cases} \int_{T_{ref}}^{T} C_p dT, & T < T_s \\ \int_{T_{ref}}^{T_s} C_p dT + \beta \Delta H, & T_s < T < T_l \\ \int_{T_{ref}}^{T_s} C_p dT + \Delta H + \int_{T_l}^{T} C_p dT, & T > T_l \end{cases} \qquad (6)$$

Where k and $C_P$ are thermal conductivity and specific heat capacity of the PCM respectively. Enthalpy h, defined by Eq. (6), is enthalpy of PCM at different states, where, $T_{ref}$ is reference temperature and the enthalpy equals to zero at this temperature. $\Delta H$ is fusion latent heat. $T_s$ and $T_l$ are the temperatures where phase change starts and ends up, respectively. Eq. (6) indicates that, in the mushy region, if $T_s = T_l = T_m$, then $T = T_m$ (present study); However, if $T_s \neq T_l$, T can be calculated by using the following equation [7]:

$$T = \beta(T_l - T_s) + T_s \qquad (7)$$

## DESIGN AND SIMULATION MODEL

**Simulation Model:** Figure 1 shows the geometry of the studied tube in the tube (shell and tube) heat exchanger with single and double pass (parallel and counter flow) arrangements. The geometry contains two concentric pipes of 14 and 60 mm diameters in single pass (figure 1A) and three concentric tubes with diameter of 14 mm, 60 mm and 70 mm for double pass (figure 1B). The length of the heat exchanger is 500 mm. the thickness of the inner tube, which are made of copper is, 1 mm. The HTF flows through the inner tube and the PCM is placed in the space between the inner and outer pipes.

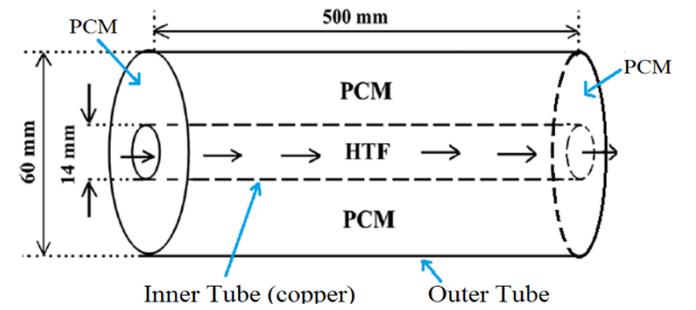

Figure 1A: Geometry of the physical model with Single Pass

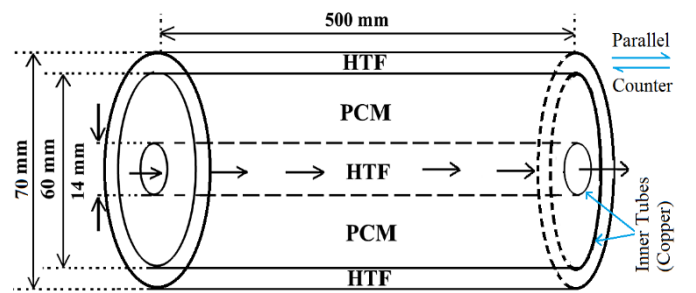

Figure 1B: Geometry of the physical model with Double Pass with Parallel and Counter flow arrangement

The Energy, Turbulence (standard k-ε) and Melting and solidification module of ANSYS fluent 18.1 were used. As density and other properties change with temperature, Boussinesq approximation with the average value of density is considered for density variation. An average value was also used for specific heat and thermal conductivity. This model treats density as a constant value in all solved equations, except for the buoyancy term in the momentum equation:

$$(\rho - \rho_0)g \approx -\rho_0 \beta'(T - T_0)g \qquad (8)$$

Where $\rho_0$ and $T_0$ are reference density and temperature and $\beta'$ is the thermal expansion coefficient. Eq. (8) is obtained by using the Boussinesq approximation $\rho = \rho_0(1 - \beta'\Delta T)$ to eliminate ρ from the buoyancy term. This approximation is accurate as long as changes in actual density are small; specifically, the Boussinesq approximation is valid when $\beta'(T - T_0) \ll 1$. The Boussinesq model should not be used if the temperature differences in the domain are large [7].

In order to discretize the energy and momentum equations the QUICK differentiating scheme is implemented. The pressure equation has been countered using the PRESTO scheme. In order to achieve a stable solution, under relaxation factors are considered which are 0.3, 0.6, 1 and 0.9 respectively for pressure, velocity, energy and volumetric liquid fraction. The convergence tolerances for the continuity, momentum and energy equations are $10^{-5}$, $10^{-5}$ and $10^{-6}$ [4].

**Phase Change Material:** Water is used as the HTF fluid and savE® OM-42 (make: PLUSS™) is used as PCM. The PCM properties [8] are given in table 1.

Table 1: Properties of phase change Material [8]

| Property | Value |
|---|---|
| Base Material | Organic |
| Melting Temp (°C) | 44 |
| Freezing Temp (°C) | 43 |
| Latent Heat (kJ/kg) | 199 |
| Liquid Density (kg/m$^3$) | 863 |
| Solid Density (kg/m$^3$) | 903 |
| Liquid Specific Heat (kJ/kg-K) | 2.78 |
| Solid Specific Heat (kJ/kg-K) | 2.71 |
| Liquid Thermal Conductivity (W/m-K) | 0.1 |
| Solid Thermal Conductivity (W/m-K) | 0.19 |
| Dynamic Viscosity [kg/m-s] | 0.023 |
| Thermal expansion coefficient [1/K] | 0.0006 |

**Boundary Conditions**: The boundary conditions used of the simulations are given in table 2.

Table 2: Boundary conditions

| Description | Boundary Condition | Input Conditions | |
|---|---|---|---|
| | | Thermal | Flow |
| Hot Water Inlet | Mass flow inlet | 338K<br>333K<br>328K | 0.1 kg/s<br>0.2 kg/s<br>0.3 kg/s |
| Hot Water Outlet | Pressure outlet | - | Atm. |
| PCM (Left) | Wall | - | - |
| PCM (Right) | Wall | - | - |
| Outer Walls | wall | Adiabatic | - |
| Inner Walls (copper) | wall | Shell Conduction | - |

The numerical simulations were performed in a transient manner with a time step of 0.5 s during charging and discharging. The melt fraction and temperature contours are also analysed to predict the heat transfer rate through the PCM. Conduction and natural convection are the prime modes of heat transfer.

## VALIDATION

Geometry presented by Kazemi et al. [4] was used to validate the results. Figure 1A shows the same geometry which was replicated and simulated in Ansys Fluent 18.1. During simulations, the results of the simple model (without fins) were reproduced with same boundary conditions (60 °C inlet temperature with 0.01 kg/s mass flow rate) with our approach by tacking same PCM RT35 properties as used in Kazemi et al. work [4]. The liquid fraction at a various time interval during the charging process was compared as shown in fig. 2.

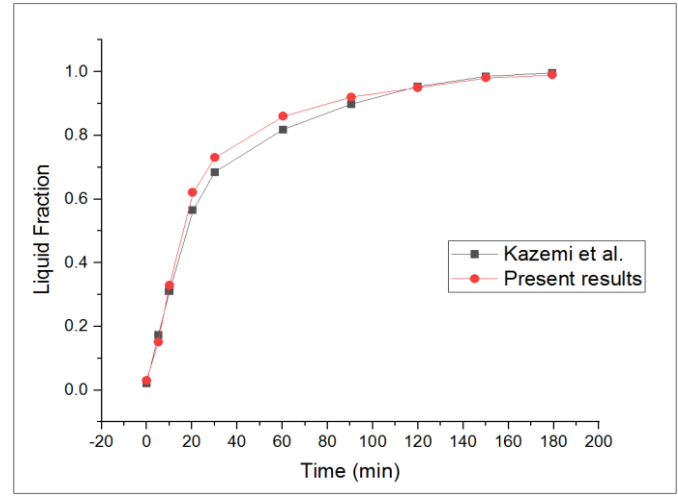

Figure 2: Validation of results (Liquid fraction versus time)

The figure 2 shows that the liquid fractions at different times of the current simulations match well with that of Kazemi et al. [4]. Hence, it is assumed that the same physics can be used for further numerical studies.

## RESULTS AND DISCUSSION

**Charging:** The temperature and liquid fraction contours at the centre (L=250 mm) in the longitudinal direction of the single pass simulation geometry at the various times during charging processes are shown in figure 3 and 4 respectively. The inlet conditions of the HTF are maintained at 65 °C temperature and 0.1 kg/s mass flow rate. Initially, the PCM is at 27 °C. With the addition of heat through the HTF, the PCM temperature and liquid fractions vary with time.

It is observed from the figure 3 and 4 that at time 5 min, the temperature gradients are more in the lower part of the cylinder in comparison to the upper part. However, the average temperature of the upper part is higher than the lower part. The melting of PCM has started near the copper tube as the temperature of HTF is beyond the melting point. The initial heat transfer mode was conduction as the whole PCM is in solid phase initially. The complete melting of the PCM had occurred near the bottom portion of the copper tube. In the upper part the tube partial melting had happened that can be observed from the liquid fraction contour. Assuming the temperature of HTF is uniform throughout the periphery of the copper tube, more temperature change can be observed during the sensible heating, whereas the temperature remains close to the constant value during the phase change. This is the reason for larger average temperature in the upper part in comparison to the lower part.

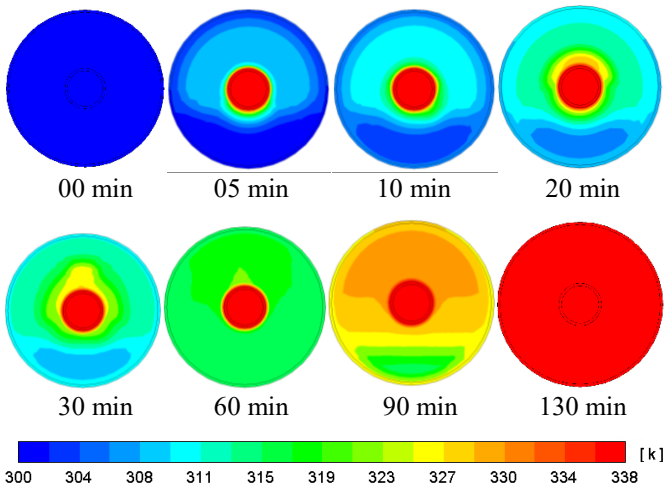

Figure 3: Temperature contours at the mid-plane with time during charging process of single pass geometry

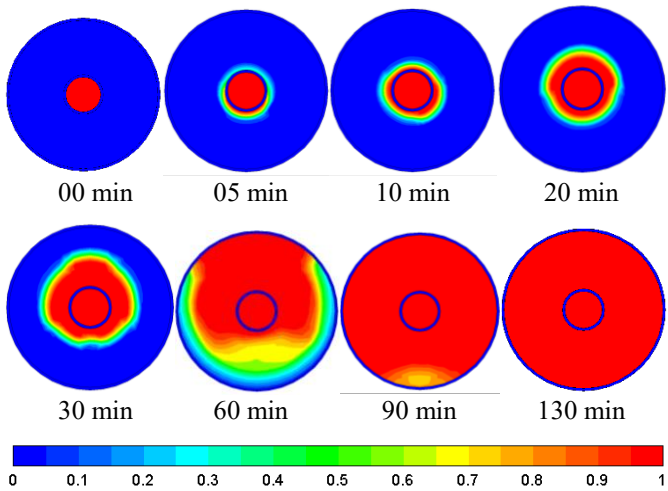

Figure 4: Liquid fraction at the mid-plane with time during charging process of single pass geometry

As the time increases to 10 min, more melting had happened near the tube and it can be seen that the temperature and liquid fraction contours are more or less uniform close to tube. As melting happens close to tube, most of the PCM away from the tube is in solid form. The quantity of the liquid fraction is small and as a result, the heat transfer is mainly due to conduction, though small convection may be present. At 20 min, more melting had happened in the upper region. The trend of the higher average temperature on the upper half continues even after 20 minutes of time. The increase in the liquid fraction in the upper half may be attributed to the convection of the fluid. In the bottom portion, the liquid fraction is less in comparison to the upper portion, and the heat transfer in the bottom portion is due to conduction only. The same trend is observed at 30 min, which is not shown in the figure. The heat transfer in the solid phase is due to conduction where the thermal conductivity of the PCM is quite low and this is responsible for the slow melting in the lower portion. At 60 min, the temperature of the whole domain approaches the melting point. At this state, the whole PCM is either in solid phase or liquid phase with uniform temperature throughout. From the liquid fraction contours, we can see that majority of the PCM in the upper half is melted, with some solid fractions close to the walls. In the lower half, the melting is not complete, which is evidenced by some solid fraction at the bottom. The reason for almost complete melting in the upper portion is attributed to the convection which is due to decrease in the density of the PCM after melting. At 90 min, the upper portion of the PCM is at temperatures above the melting point. As a result, the energy storage is in the form of sensible storage. However, in the bottom portion, the temperatures are close to the melting point. Except for a small portion closer to the lower wall of the PCM casing, the whole PCM is melted. Heat addition after melting is stored in the form of sensible heating, hence, the energy storage capacity decreases in comparison to the latent heat storage. The HTF adds heat till 130 minutes. At 130 min the temperature of whole PCM reached approximately to 65 °C, which is due to sensible heating after the whole PCM reached beyond melting temperature.

**Discharging:** The figure 5 shows the liquid fraction at the mid-plane with time during discharging process (energy extraction) of single pass geometry. The starting point of the discharging is the end conditions of the melting, that is discussed in the previous section. For the discussion purpose, we assume the leftover condition of the charging (65 °C), as the starting point of discharging and time at this moment is taken a 0 min. For getting the total time from the beginning of charging, we can add the discharge time to 130 min. The inlet conditions of the HTF are maintained at 20 °C temperature and 0.1 kg/s mass flow rate throughout the simulation. Initially, the whole PCM is at 65 °C.

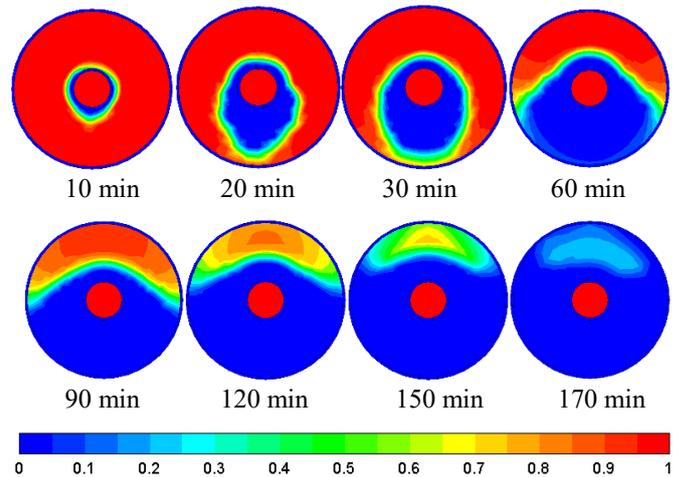

Figure 5: Liquid fraction at the mid plane with time during discharging process of single pass geometry

During discharging the phenomenon is just opposite to charging process. The solidification starts from bottom of tube and it propagate to downward direction first. Since solidification process is conduction dominant and PCM suffers with low thermal conductivity, solidification (discharging) is very slow process and discharging takes almost double time as compared to charging. The liquid fraction profile for the discharging process is shown in figure 5. At the beginning (10 min), the solidification started at the bottom of the HTF tube. As the time increases, more solidification is observed in the bottom portion of the PCM. This is due to the convection. With further discharging, we can observe that whole of the bottom portion is solidified. In the top portion, PCM is still in the liquid form. We can observe that even after 170 minutes, some PCM is in liquid form at the top. In comparison to the charging, where the PCM has melted completely by 130 minutes, the discharging takes more time. This difference might be because of the conduction - convection dominance during discharge while convection is dominant during the charging. The temperature difference between the melting point of PCM and HTF during charging and discharging are almost same (22 °C during charging and 23 °C during discharging). Numerical simulation results for discharging clearly indicate that the initial discharging was quite fast then it became slower with time. We can observe that 50% of the solidification happened during first 60 minutes. Similarly, 50% of the melting happened during the first 60 minutes of charging. From this, we can expect that for the fast latent heat storage and extraction, partial charging and discharging gives better results as compared with fully charging and discharging. Depending on requirement, the mode of charging and discharging (fast or slow or combination) need to be explored.

The results of discharging process shown in Figure 5 are asymmetric which is generally due to convection heat transfer. This is due to the fact that, although it is conduction dominant but it is not a pure conduction case, else it also include simultaneous local convection. Added to this the thermal conductivities of liquid and solid phases are different. If there is only conduction the results would have been symmetric.

**The Effect Mass Flow Rate and Inlet Temperature:**

The effect of HTF mass flow rate on the liquid mass fraction of PCM is presented in figure 6. It is observed that the mass flow rate of HTF has a very minimal effect on the melting phenomena. However, all the curves show a similar trend with respect to the liquid fraction.

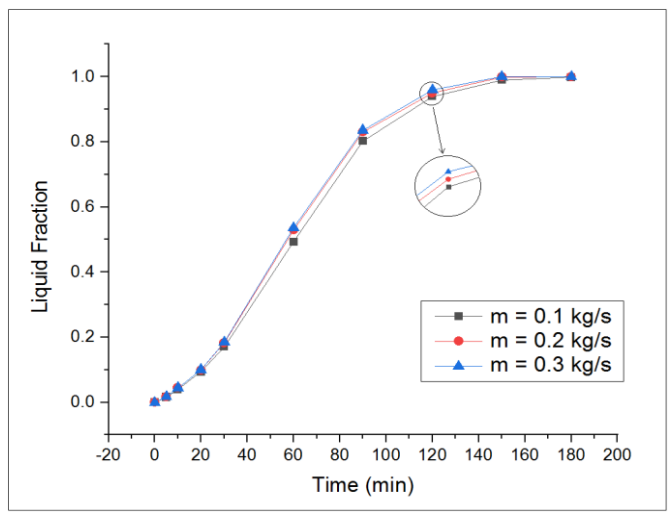

Figure 6: Effect of mass flow rate in melting fraction

In all cases, it can be seen that the melting starts slowly, as evidenced by the smaller slope of the curve, followed by a constant slope, followed by a flat curve. This may be explained by considering the fact that initially, the heat transfer mechanism is by conduction and the melting process is slow. As the time proceeds, the convection heat transfer occurs, and the rate of melting increases. This is evidenced by the increase in the slope. As the time proceeds, most of the PCM is in the liquid form, with small portions of the PCM in the solid form at the bottom. As the convection in the bottom portion is not dominant, the solid takes a long time to melt. This can be observed by having a flat natured curve. As the mass flow rate increases the heat transfer become fast, but the effect of mass flow rate is limited for certain range of increment. An optimization study may be carried in future to find the threshold mass flow rate below which the mass flow rate effect on the heat transfer is significant.

The effect of HTF inlet temperature is very considerable on the charging phenomenon (melting). Figure 7 shows the melting fraction of PCM with time, with the inlet temperature of HTF. It is observed that the melting phenomena are hugely affected by the inlet temperature of HTF. At lower HTF inlet temperature (55 °C), the melting process is fairly uniform with time. Liquid fraction slowly starts and maintains a uniform slope till the whole (or close to complete) PCM has melted. At 60 °C HTF inlet temperature, the melting process is fast, in comparison to 55 °C HTF inlet temperature, till a liquid fraction of 0.8. Beyond this liquid fraction, the melting slows down due to the stagnant region in the bottom of the PCM cylinder. With 65 °C HTF inlet temperature, the melting process is even faster. However, most of the melting happened by 90 min in this case. Beyond 90 min, the heat transfer happens to the liquid PCM and some to the solid PCM at the bottom. The energy storage is mostly in the form of sensible heating after 90 min. It can be observed from the figure that majority of melting (90% in the case of 65 °C HTF inlet temperature and 80% in the case of 60 °C HTF inlet temperatures) takes place in 50 percent of the time. In the rest of the time, the temperature rise of PCM takes place rather than melting.

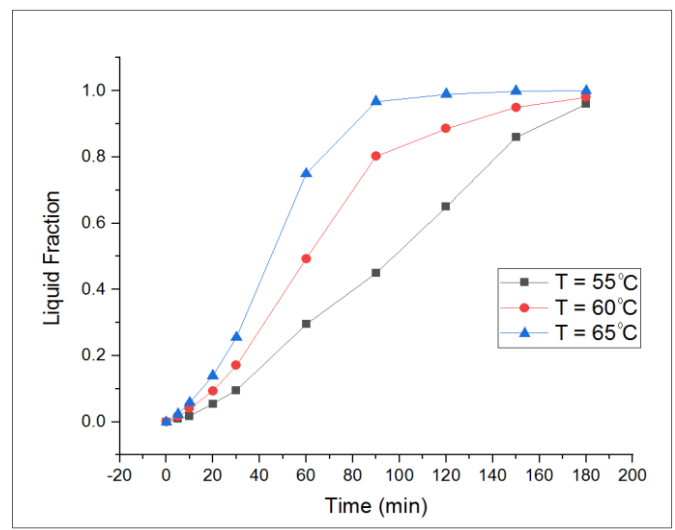

Figure 7: Effect of hot water inlet temperature variation in melting fraction

The Liquid Fraction and average temperature variation with time, for one complete cycle of charging and discharging of PCM is shown in figure 8 and 9.

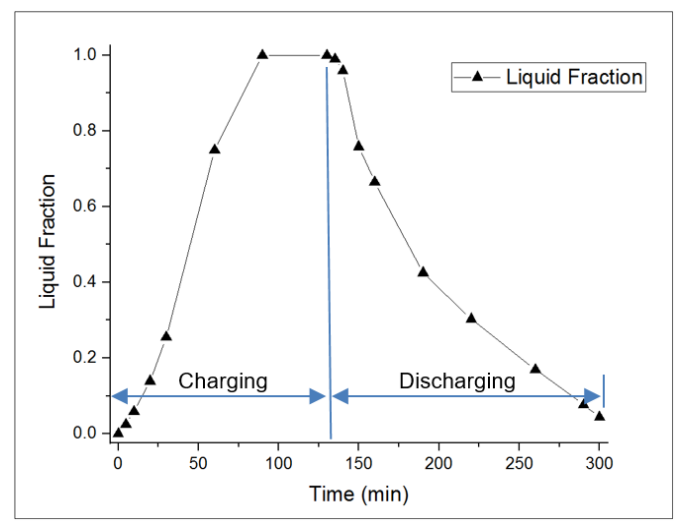

Figure 8: Liquid Fraction v/s Time Graph for one complete cycle of charging and discharging of PCM

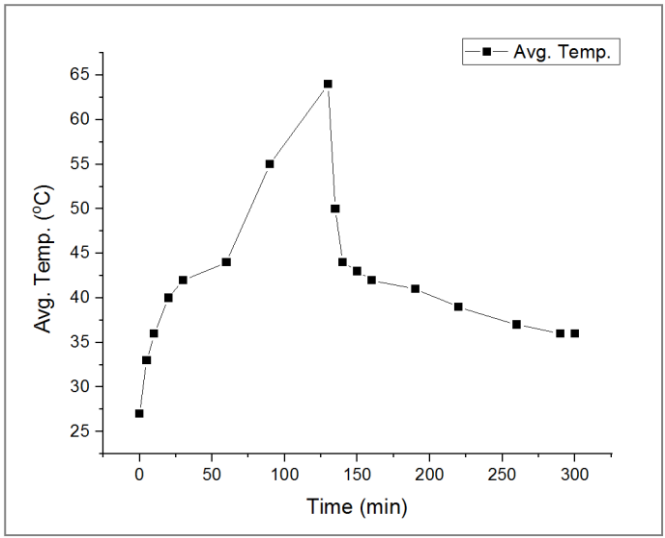

Figure 9: Avg. Temperature v/s Time Graph for one complete cycle of charging and discharging of PCM

From the figures, it is clear that charging takes less time (100 min) as compared to discharging (170 min). After 100 minutes, the PCM is just heated up in the liquid form. From figure 9 it is clear that once PCM is fully charged, temperature of PCM increasing due to sensible heating, which are not recommended for many applications where constant temperature is required. During the discharging, we can see that the temperature falls sharply initially due to sensible cooling. Once the PCM reaches the solidification temperature, we can observe that the temperature decreases slowly.

This is because the convection dominates the charging and conduction dominates the discharging. Hence for the shorter charging and discharging time as well as for efficient design, further simulations are required with various HTF mass flow rates and inlet temperatures.

**Double Pass Latent Heat Energy Storage System:**

Due to high specific heat, some HTFs like water have sufficient energy to further charge the PCM. There is a need to study the impact of different heat exchanger configurations on the charging and discharging of PCM.

Double pass (Triple Tube) Latent Heat Energy Storage System (LHESS) is one of the possibility by which we can decreases the charging and discharging time considerably. As heat transfer areas increase, the heat transfer rates also increase and one may expect uniform liquid fractions and temperatures and the problems of solid and liquid regions of PCM at bottom and top are also significantly reduced. Since melting (convection current) starts from both side of PCM, the charging rate become faster and more uniform.

The effect of mass flow rate and inlet conditions have similar trend as in single pass, with significantly reduced charging time. These results are not shown here, as they merely present any new insight. It may vary with different inner and outer pipe diameter ratio.

The results with parallel and counter flow arrangements, water as HTF, the performance is quite similar with some edge to counter flow. The further analysis of double pass configuration is still in progress. Some other configurations like helical coil etc. also need to be analyze for more efficient design of LHESS.

## Conclusions

In the current study, numerical simulations were carried out to find the charging and discharging time of the PCM cylinder. Numerical simulations were performed with various inlet mass flow conditions and inlet temperatures with single and double pass (counter and parallel) arrangements of HTF. The following conclusions are drown from it:

➢ It was observed that the HTF inlet temperature has the greater impact on the charging and discharging as compared to that of the mass flow rate.
➢ It is also observed that more uniform charging happens at lower HTF inlet temperature.
➢ Partial charging and discharging is more effective as 80-90% charging and discharging takes around 50% time.
➢ With double pass, charging rate become fast and more uniform as compared to single pass.

Findings from present simulations can be used for better design of latent heat energy storage system.

## Acknowledgments

Authors acknowledge the financial support provided by SERB, DST through grant ECR/2015/00526, and project titled 'Solar Energy Storage Using Phase Change Materials for Space Heating Applications'.